\documentclass[twocolumn, english]{article}
\usepackage[affil-it]{authblk}
\usepackage{refstyle}
\usepackage[T1]{fontenc}
\usepackage[latin9]{inputenc}
\usepackage{textcomp}
\usepackage{amsmath}
\usepackage{amssymb}
\usepackage{mathrsfs}
\usepackage{esint}
\usepackage{babel}

\usepackage{color,hyperref}
    \catcode`\_=11\relax
    \newcommand\email[1]{\_email #1\q_nil}
    \def\_email#1@#2\q_nil{%
      \href{mailto:#1@#2}{{\emailfont #1\emailampersat #2}}
    }
    \newcommand\emailfont{\sffamily}
    \newcommand\emailampersat{{\color{red}\small@}}
    \catcode`\_=8\relax

\makeatletter

\RS@ifundefined{subref}
  {\def\RSsubtxt{section~}\newref{sub}{name = \RSsubtxt}}
  {}
\RS@ifundefined{thmref}
  {\def\RSthmtxt{theorem~}\newref{thm}{name = \RSthmtxt}}
  
  {}
\RS@ifundefined{lemref}
  {\def\RSlemtxt{lemma~}\newref{lem}{name = \RSlemtxt}}
  {}

\usepackage{amsmath}

\makeatother
\begin{document}
\title{Separability and entanglement in classical eigenfunctions as a criterion for Hamiltonian chaos}
\author{A. D. Berm\'udez Manjarres}
\affil{\footnotesize Universidad Distrital Francisco Jos\'e de Caldas\\ Cra 7 No. 40B-53, Bogot\'a, Colombia\\ \email{ad.bermudez168@uniandes.edu.co}}
\twocolumn[
  \begin{@twocolumnfalse}

\maketitle

\centering
    \begin{minipage}{.9\textwidth}
\begin{abstract}
We study the eigenfunctions of the classical Liouville operator and investigate the conditions they must obey to be separable as a
product state. We point out that the conditions for separability are equivalent to the requirements of Liouville's integrability theorem; that is, the eigenfunctions are separable if and only if the system is integrable. On the other hand, if the classical system is not integrable, then the eigenfunctions are entangled in all canonical coordinates. This links the classical notions of chaos and integrability with mathematical concepts usually restricted to quantum mechanics. 
\end{abstract}

\hspace{2cm}
    \end{minipage}
\end{@twocolumnfalse}
]

\section{Introduction}

The defining signature of chaos in classical mechanics is the exponential separation of neighboring trajectories, as measured by the Lyapunov exponents \cite{Lyapunov}. This trajectory-based approach to chaos cannot be used in quantum mechanics due to the Heisenberg uncertainty principle. Hence, quantum chaos is usually investigated using operator and Hilbert space methods \cite{Qchaos,Qchaos2}. 

Since the classical trajectory approach does not have a quantum analog, the investigations of quantum chaos in phase space are based on the (quasi)probability distribution given by the Wigner function \cite{Qchaos3,Qchaos4}. In concordance, it has been argued, for example by Ballentine and Zibin \cite{Rchaos}, that a proper comparison between quantum and classical chaos requires a study of the dynamics of classical probability distribution given by the Liouville equation. We can find several works that use this approach, see for example \cite{Rchaos2,Rchaos3,Rchaos4,Rchaos5,Rchaos6,Rchaos7} and references therein. 

Following the same logic, we go a step further and investigate integrability and chaos in Hamiltonian systems using \emph{classical} complex-valued wavefunctions and other mathematical concepts originally developed in quantum mechanics. Our use of complex functions is not arbitrary because they arise naturally when studying the spectrum of differential operators from classical mechanics.

Our interest is twofold. First, from a purely classical point of view, Hamiltonian
chaos is still a hard and puzzling subject. Any novel insight that
can be given into the matter is of value. 

Second, studying classical chaos using concepts from quantum mechanics could inspire novel approaches
to quantum chaos (we comment further about this point in the Appendix). In this regard, it is of particular interest that here we give a criterion for chaos that is independent of the exponential divergence of the trajectories, and, contrary to \cite{Rchaos7}, seems to be independent of Lyapunov exponents. 

It is important to remark here that we will not use any classical or semi-classical limit. Instead, we will employ the natural Hilbert space structure that arises in classical mechanics as first studied by Koopman
and von Neumann \cite{koopman,von neumann,mauro,Prigogine}
(hereafter referred to as the Koopman-von Neumann (KvN) theory). The KvN theory and
related formalisms have received renewed
interest in recent years. A primary reason for this is that writing Hamiltonian dynamics as a Hilbert space theory
allows us to compare and contrast classical and quantum mechanics. 

We can also mention that techniques originally developed for quantum mechanics have been recently applied to purely classical systems \cite{tool 1,tool 2,tool 3,tool 4,tool 5,tool 6,tool 7,tool 8,adiabatic driving,tool 9}. 

A third reason for the interest in the classical Hilbert space and classical wavefunctions is the development of quantum-classical hybrid models \cite{hybrid1,hybrid2,hybrid 3,reginatto}.

Finally, we mention that the study of the classical wavefunctions has given insights into the classical and semi-classical limits of quantum
mechanics \cite{joseph,wilkie,wilkie 2,bondar 1,bondar 2}. 

We want to highlight the recent use of the KvN theory to study the
transition into chaos in Hamiltonian systems \cite{adiabatic driving}.
It was previously suggested that adiabatic transformations could offer
a common framework to study both classical and quantum chaos \cite{polkovnikov},
and it was shown in \cite{adiabatic driving} that the KvN formalism
is well-equipped to explore these transformations. As a result, it
was noted in \cite{adiabatic driving} that the transition into chaos
shares similarities with a quantum phase transition, thus linking
two seemingly unrelated concepts, and advancing the idea that quantum and classical chaos
can be understood on the same footing.

Here, we will show a relationship between the classical ideas of integrability
and chaos with the mathematical notions of separability and entanglement of wavefunctions, two concepts that first appeared in the context of quantum mechanics. To do so, we will investigate the eigenfunctions of the time-evolution operator of the KvN theory, the so-called Liouvillian
operator. We focus on the requirements the Liouville eigenfunctions
must obey to be written as a product state, and how this is related
to integrability and chaos in Hamiltonian systems. We will show that
the conditions for separability of the Liouville eigenfunctions are
equivalent to the requirements of Liouville's integrability theorem
\cite{Arnold}, i.e., the eigenfunctions are separable if and only
if the system is integrable. On the other hand, the eigenfunctions of a chaotic system are entangled in all canonical coordinates. 

Specifically, the situation is as follows: let $\mathbf{z}$ be any
arbitrary set of canonical variables, then the odds are that the eigenfunctions
of the Liouville operator \emph{do not} factorize as a product state
$\psi(\mathbf{z})\neq\psi_{1}(z_{1})...\psi_{2N}(z_{2N})$. However,
if (and only if) the system is integrable in the Liouville sense, then there exists a continuous canonical transformation $\mathbf{z}\rightarrow\mathbf{Z}=\mathbf{Z}(\mathbf{z})$
such that the eigenfunctions are a product state when written
in terms of the new variables i.e., $\psi(\mathbf{Z})=\psi_{1}(Z_{1})...\psi_{2N}(Z_{2N})$.

The situation described above is analogous to the quantum case. As emphasized in \cite{single}, it is well known that a quantum wavefunction can be separable in one set of coordinates but entangled in some other coordinates. For example, the ground state
of an isolated hydrogen atom is non-factorizable and hence entangled when written in terms of the proton and electron coordinates \cite{hydrogen}.
However, the same state can be factorized when written in the coordinates
of the center of mass and the relative motion. The fundamental point
is that entanglement is not an absolute property of a quantum state,
but it is always a property of the state relative to a given set of
subsystems. In the classical case, the state is given by the Liouville eigenfunctions, and the role of subsystems is taken by the canonical coordinates used. 

It is important to emphasize that we are dealing with mathematical analogies between the quantum and classical wavefunctions. When we say that a classical wavefunction is entangled in some canonical coordinates, we mean that such wavefunction does not factorize in those coordinates, i.e., it is a description of a mathematical object. We are not implying here that classical mechanics allows for quantum behaviours like the many phenomena due to the entanglement between quantum particles.

In the next section, we give a short review of the main elements
of the KvN theory. We present only the essential elements of the theory
as we will not need more for our purposes here (we refer to \cite{mauro} for a more complete presentation). 

In section 3, we investigate the problem of the factorizability of Liouville eigenfunctions and their relationship with the Liouville integrability
theorem. The main results of the paper are given in this section,
with section 3.2 having the novel approach to Liouville integrability. 

\section{A quick overview of the KvN theory}

The KvN formulation of classical (statistical) mechanics starts by rewriting the Liouville equation
\begin{equation}
\frac{\partial\rho}{\partial t}+\left\{ H,\rho\right\} =0
\end{equation}
in the Schr\"{o}dinger-like form

\begin{equation}
i\frac{\partial\psi}{\partial t}=\hat{\vartheta}_{H}\psi,\label{KvN}
\end{equation}
where $\left\{ ,\right\} $ is the Poisson bracket, the so-called
Liouvillian operator is defined by $\hat{\vartheta}_{H}=-i\left\{ ,H\right\} $,
and the wavefunction $\psi$ is related to the probability density
by the Born rule $\rho=\left|\psi\right|^{2}$. This simple trick allows us to use well-known mathematics from quantum mechanics in
a classical context. 

The wave functions belong to the classical Hilbert space of square-integrable
functions over the phase space $T^{*}Q$, where the inner product
is given by

\begin{equation}
\left\langle \varphi\right|\left.\psi\right\rangle =\int_{T^{*}Q}d\mathbf{z}\,\varphi^{*}\psi.
\end{equation}
However, just as in quantum mechanics, the KvN theory requires the
use of non-normalizable wavefunctions as they appear as eigenfunctions
of some of the important operators. This is the case of the Liouvillian, for we will see that its eigenfunctions are distribution-valued.

In the KvN theory, there are two kinds of self-adjoint operators of
relevance that are constructed from phase-space functions. For any
continuous function $f(\mathbf{z})$ we can define a multiplication
operator $\hat{f}\psi(\mathbf{z})=f(\mathbf{z})\psi(\mathbf{z})$.
The multiplication operators have the role of the measurable quantities, as the classical mean values are given by

\begin{equation}
\left\langle f\right\rangle =\int_{T^{*}Q}d\mathbf{z}\,\psi^{*}\hat{f}\psi=\int_{T^{*}Q}d\mathbf{z}\,f\rho.
\end{equation}

The second kind of operators associated with $f$ are the vector fields
$\hat{\vartheta}_{f}=-i\left\{ ,f\right\} $. These differential operators
are the generators of unitary transformations (or canonical transformations
from the classical Hamiltonian point of view). For example, the Liouvillian
is the generator of time evolution
\begin{equation}
\psi(\mathbf{z},t)=\exp\left[-i\hat{\vartheta}_{H}t\right]\psi(\mathbf{z},0).
\end{equation}

It is one of the unusual quirks of the KvN theory that, unlike quantum mechanics, observables, and generators are given by different
operators \footnote{Using van Hove's work on the unitary representation of the group of
canonical transformations \cite{hove}, it is possible to give a classical Hilbert space theory where the observables and the generators coincide
\cite{reginatto}. We will not use this more sophisticated formalism
here, though.}. 

The algebra of interest for the KvN operators is given by the following
equations \cite{pauri}: 

\begin{align}
\hat{\vartheta}_{fg} & =\hat{f}\hat{\vartheta}_{g}+\hat{g}\hat{\vartheta}_{f},\\{}
[\hat{\vartheta}_{f},\hat{g}] & =[\hat{f},\hat{\vartheta}_{g}]=i\{f,g\},\label{commute}\\
\hat{\vartheta}_{\{f,g\}} & =-i[\hat{\vartheta}_{f},\hat{\vartheta}_{g}],\\{}
[\hat{f},\hat{g}] & =0.
\end{align}

\section{Liouville eigenfunctions}

For autonomous Hamiltonian systems, we are interested in the eigenvalue
problem 

\begin{align}
\hat{\vartheta}_{H}\psi(\mathbf{z}) & =\lambda\psi(\mathbf{z}).\nonumber \\
\hat{H}\psi(\mathbf{z}) & =E\psi(\mathbf{z}).\label{eigenfunction}
\end{align}
The problem of finding the Liouville eigenfunctions is highly non-trivial in most cases, but sometimes there is an analytical solution. For example, consider the Hamiltonian for the harmonic oscillator

\[
H=\frac{p^{2}}{2m}+\frac{m\omega^{2}}{2}q^{2}.
\]
The Liouville eigenfunctions for this system are given by \cite{jaffe}

\begin{equation}
\psi_{E,n}(q,p)=\frac{\omega}{2\pi}\delta(\frac{p^{2}}{2m}+\frac{m\omega^{2}}{2}q^{2}-E)\,\left[\frac{p+im\omega q}{p-im\omega q}\right]^{n},\label{eigenpsi}
\end{equation}
where $n$ is an integer, and the eigenvalues are $\lambda_{n}=n\omega$.
Notice that these eigenfunctions do not factorize as a product state
\[
\psi_{E,n}(q,p)\neq f(q)g(p),
\]
thus, the canonical coordinates $q$ and $p$ are entangled, and their statistical properties are correlated. However, we will see in the next section that for the harmonic oscillator (and other integrable
systems), it is possible to make a canonical transformation such that the eigenfunctions are factorizable when written in terms of the new
canonical variables.

\subsection{Integrability implies separability}

An autonomous Hamiltonian system of $N$ degrees of freedom is said
to be Liouville integrable if there exist $N-1$ additional independent isolating
integrals of the motion $K(\mathbf{z})$ in involution\cite{Arnold},
i.e., 

\[
\{K_{i},K_{j}\}=\{K_{i},H\}=0,
\]
for all $i$ and $j$. This mean that $\hat{H},\hat{K}_{1},...,\hat{K}_{N-1}$
and $\hat{\vartheta}_{H},\hat{\vartheta}_{K_{1}},...,\hat{\vartheta}_{K_{N-1}}$
form a complete set of commuting self-adjoint operators, and we can
find common eigenfunctions for them, given by \cite{jaffe}

\begin{equation}
\psi(\mathbf{z})=\delta(E-H(\mathbf{z}))e^{i\lambda\tau(\mathbf{z})}\prod_{j=1}^{n-1}\delta(k_{j}-K_{j}(\mathbf{z}))\,e^{i\lambda_{j}\eta_{j}(\mathbf{z})}\label{psiK}
\end{equation}
where $\left\{ \tau,H\right\} =1$ and $\left\{ \eta_{i},K_{j}\right\} =\delta_{ij}$.

Moreover, if the system has compact energy levels, then there exists
a canonical transformation to angle-action variables $(\mathbf{I},\pmb{\phi})$
such that the Hamiltonian is only function of the actions $H(\mathbf{z})=H(\mathbf{I})$.
In this case, the Liouvillian becomes 

\begin{align}
\hat{\vartheta}_{H} & =\pmb{\omega}\cdot\hat{\vartheta}_{\mathbf{I}}=-i\pmb{\omega}\cdot\frac{\partial}{\partial\pmb{\phi}},\nonumber \\
\omega_{j} & =\frac{\partial H}{\partial I_{j}},
\end{align}
and the normalized eigenfunctions are given by \cite{wilkie}

\begin{equation}
\psi_{\mathbf{I}',\mathbf{k}}=\frac{1}{(2\pi)^{N/2}}\delta(\mathbf{I}'-\mathbf{I})e^{i\mathbf{k}\cdot\pmb{\phi}},\label{psiI}
\end{equation}
with eigenvalues
\[
l_{\mathbf{I}',\mathbf{k}}=\mathbf{k}\cdot\pmb{\omega}(\mathbf{I}'),
\]
where $\mathbf{k}\in\mathbb{Z}^{n}$.

We can see that the eigenfunctions (\ref{psiI}) are written as a
product of eigenfunctions of the $N$ actions $I_{i}$ and the $N$
differential operators $-i\{,I_{i}\}=i\frac{\partial}{\partial \phi_{i}}$.
We can then conclude that the angle-action variables $(\mathbf{I},\pmb{\phi})$
are statistically uncorrelated. A similar observation can be given
for the more general form (\ref{psiK}). 

\subsection{Separability implies integrability }

Let us now assume that there exists a continuous canonical transformation
$\mathbf{z}\rightarrow\mathbf{Z}=\mathbf{Z}(\mathbf{z})$ such that the eigenfunctions of the Liouvillian are separable in a domain $D\subseteq T^{*}Q$
when written in terms of $\mathbf{Z}$. This is, $\psi$ can be written
as
\begin{equation}
\psi(\mathbf{Z})=\psi_{1}(Z_{1})...\psi_{2N}(Z_{2N}).\label{psi}
\end{equation}
The continuity of the transformation guarantees that the multiplicative
operators $\mathbf{\hat{\mathbf{Z}}}$ are self-adjoint\cite{reed-simon}. 

In terms of the new variables $\mathbf{Z}$, the eigenvalue Eqs. (\ref{eigenfunction})
read

\begin{align}
\hat{H}(\hat{\mathbf{Z}})\psi_{1}(Z_{1})...\psi_{2N}(Z_{2N}) & =E\psi_{1}(Z_{1})...\psi_{2N}(Z_{2N}),\label{eigen1}\\
\hat{\vartheta}_{H}\psi_{1}(Z_{1})...\psi_{2N}(Z_{2N}) & =\lambda\psi_{1}(Z_{1})...\psi_{2N}(Z_{2N}),\label{eigen2}
\end{align}
where the Liouvillian operator can be written as

\[
\hat{\vartheta}_{H}=\sum_{i,j}\frac{\partial H}{\partial Z_{j}}J_{ij}(-i\frac{\partial}{\partial Z_{i}}),
\]
and the $J_{ij}$ are the elements of the symplectic form defined
by $\{Z_{i},Z_{j}\}=J_{ij}$.

We will now show that these eigenvalue equations imply the existence in $D$ of $N$ functionally independent constants of motion in involution. 

The proof starts by showing that the Eqs (\ref{eigen1}) and (\ref{eigen2})
together imply that each $\psi_{i}(Z_{i})$ is either an eigenfunction
of $Z_{i}$ or an eigenfunction of $-i\frac{\partial}{\partial Z_{i}}$. 

We proceed as follows: let us start with Eq. (\ref{eigen1}) and focus
on the coordinate $Z_{i}$. Since $\hat{H}(\hat{\mathbf{Z}})$ is
a function of multiplicative operators, its action on the eigenfunctions
is of the form
\[
\hat{H}(...\hat{Z}_{i}...)...\psi_{i}(Z_{i})...=H(...Z_{i}...)...\psi_{i}(Z_{i})....
\]
Hence, we conclude that $\psi_{i}(Z_{i})$ must be an eigenfunction
of $\hat{Z}_{i}$ \emph{unless }the Hamiltonian happen to be independent
of $\hat{Z}_{i}$.

On the other hand, For Eq. (\ref{eigen2}) to be satisfied, the function
$\psi_{i}(Z_{i})$ must be an eigenfunction of $\frac{\partial H}{\partial Z_{j}}J_{ij}(-i\frac{\partial}{\partial Z_{i}})$
for the value of $j$ that gives the conjugate variable to $Z_{i}$.
For example, we must have for the kth coordinate that $\psi(Q_{k})$
is eigenfunction of $\frac{\partial H}{\partial P_{k}}(-i\frac{\partial}{\partial Q_{k}})$.
Hence, each  $\psi_{k}(Z_{i})$ must be an eigenfunction of $-i\frac{\partial}{\partial Z_{i}}$ \emph{unless} the Hamiltonian happens to be independent of the conjugate variable of $Z_{i}$.

However, $\psi_{i}(Z_{i})$ cannot be an eigenfunction of both $\hat{Z}_{i}$
and $(-i\frac{\partial}{\partial Z_{i}})$ since these are non-commuting
operators. Thus, for Eqs (16) and (17) to be both true at the same time, it is required that: (1) The Hamiltonian is independent of $Z_{i}$
such that $\psi_{i}(Z_{i})$ does not need to be an eigenfunction
of $\hat{Z}_{i}$, or (2) the Hamiltonian is independent of $Z_{j}$,
the conjugate coordinate to $Z_{i}$, such that $\frac{\partial H}{\partial Z_{j}}=0$
so that the term $\frac{\partial H}{\partial Z_{j}}(-i\frac{\partial}{\partial Z_{i}})$
vanishes, and then $\psi_{i}(Z_{i})$ does not need to be an eigenfunction
of $-i\frac{\partial}{\partial Z_{i}}$ \footnote{Without loss of generality, we can ignore the case where the Hamiltonian
is independent of both members of a pair of conjugate canonical coordinates.
A situation like this implies that the systems have one less degree
of freedom than originally thought, and the dynamics is restricted
to a lower-dimensional manifold. We can then completely ignore the spurious coordinates, and our results hold for the reduced system.}.

The above means that, for a given pair of conjugate variables $Q_{i}$
and $P_{i}$, the Hamiltonian can only depend on $Q_{i}$ or $P_{i}$ but not both. Thus, at least one member of each pair of the new
canonical coordinates is absent from the Hamiltonian, and, from Hamilton's equations, the other pair member is a constant of the motion.
Moreover, these constants of motion are independent and in involution because
they are unpaired canonical coordinates. $\blacksquare$

\section{Discussion}

We studied the eigenfunctions of the (classical) Liouville operator, and we found that the factorizability of these eigenfunctions by a continuous canonical transformation is a criterion for the Liouville integrability of the system. On the other hand, the absence of such a canonical transformation is a signature of chaos. 

Our approach to classical chaos is based on the Hilbert space formulation of classical mechanics. We argue that such an approach is valuable because it might open the door for a unifying framework to study classical and quantum chaos. Often, it is mentioned that quantum and classical chaos are strikingly different because the Schr\"{o}dinger equation is linear, and classical chaos arises due to non-linearities in the Hamilton equations. However, to properly compare the two theories, we must first bring them into a common ground, as the one given by the KvN theory.  Like the Schr\"{o}dinger equation, the KvN equation is linear, yet it allows for chaotic dynamics. 

The results shown in this paper might open new lines of research by showing
that concepts usually limited to the quantum realm can be of value in classical scenarios.

Reciprocally, using Hilbert space methods to reformulate Hamiltonian chaos might open new paths that lead to new understandings of quantum
chaos. Based on the results we got in the main text, we suggest in the appendix a possible line of investigation using quantum phase-space
wavefunctions.

\subsection*{Appendix: On the Separability of the quantum phase-space wavefunctions }

It is widely known that quantum mechanics can be formulated in phase
space via the Wigner functions \cite{wigner}. Less known is that we can define an entire family of quantum phase-space wave functions
\cite{qphasespace,qphasespace2,qphasespace3}, and that these formalisms
are equivalent to standard quantum mechanics. 

In the classical limit, these quantum wavefunctions go to KvN (and
related) wavefunctions \cite{bondar 2}. Given such a close relationship,
it is worth asking if we can use the entanglement of the quantum phase-space
eigenfunctions as a criterion to define quantum chaos, just as we
did for the classical case. I cannot answer this question, but here I give an example of quantum eigenfunctions that can
be factorized by a canonical transformation. 

Consider the eigenvalue problem for the quantum Harmonic oscillator

\begin{align}
\hat{H}\psi & =\left(\frac{\hat{P}^{2}}{2m}+\frac{m\omega^{2}}{2}\hat{Q}^{2}\right)\psi=E\psi,\label{harmonic}
\end{align}
where the position and momentum operators obey the Heisenberg commutation
relation $[\hat{Q},\hat{P}]=i\hbar$. A phase-space representation
of the wavefunctions can be given by the use of the Bopp representation

\begin{equation}
\hat{Q}=q+\frac{i\hbar}{2}\frac{\partial}{\partial p},\;\hat{P}=p-\frac{i\hbar}{2}\frac{\partial}{\partial q},
\end{equation}
and this results in a so-called symplectic (time-independent) Schr\"{o}dinger
equation \cite{symplectic}. The solution to the above eigenvalue problem in phase space is known \cite{symplectic}. The eigenvalues
of (\ref{harmonic}) are $E_{n}=\hbar\omega(n+1/2)$, and the phase space eigenfunctions are given by
\begin{align}
\psi_{0}(q,p) & =\sqrt{\frac{e}{\pi\hbar}}\exp\left[-\frac{1}{\hbar\omega}\left(p^{2}/m+m\omega^{2}q^{2}\right)\right],\nonumber \\
\psi_{n}(q,p)= & \frac{3^{n}}{\sqrt{n}}\left[\sqrt{\frac{e}{\pi\hbar}}\left(q-\frac{ip}{m\omega}\right)\right]^{n}\nonumber \\
 & \times\exp\left[-\frac{1}{\hbar\omega}\left(p^{2}/m+m\omega^{2}q^{2}\right)\right],\;n\geq1.\label{psixyz}
\end{align}

For $n\geq1$ we can see that the wavefunctions are entangled $\psi_{n}(q,p)\neq f(q)g(p)$.
However, just as for the classical harmonic oscillator, the wavefunctions can be factorized
by a transformation to angle-action variables defined by

\begin{align}
q & =\sqrt{\frac{2I}{m\omega}}\sin\phi,\nonumber \\
p & =\sqrt{2m\omega I}\cos\phi.
\end{align}
The wavefunctions in terms of $(I,\phi)$ become

\begin{align}
\psi_{0}(I,\phi) & =\sqrt{\frac{e}{\pi\hbar}}e^{-2I/\hbar},\nonumber \\
\psi_{n}(I,\phi) & =\frac{3^{n}}{\sqrt{n}}\left(\frac{2eI}{\pi\hbar m\omega}\right)^{n/2}e^{-2I/\hbar}e^{in(\phi-\pi/2)},\;n\geq1.
\end{align}

Just as in the classical case, we can see that the angle variable appears only in the phase of the eigenfunction.

The above generalizes trivially to the case of a chain of interacting harmonic oscillators via a transformation to normal coordinates.

It is an open problem to determine in general if (or when) the eigenfunctions
of the symplectic Schr\"{o}dinger equation corresponding to classical integrable systems can also be factorized by a transformation to the
classical angle-action variables or any other suitable canonical transformation.

\end{document}